\documentclass[a4paper,12pt]{article}

\usepackage{indentfirst}

\usepackage{amsfonts,amsmath,amssymb,bm,a4wide,graphicx,cite,makeidx,multicol}

\begin{document}

\title{Reply to hep-ph/0605114}

\author{Alexander Grigoriev
   \\
   \small {\it Skobeltsyn Institute of Nuclear Physics, Moscow State University, 119992 Moscow,  Russia }
   \\
        Andrey Lobanov , \
        Alexander Studenikin
   \\
   \small {\it Department of Theoretical Physics, Moscow State University, 119992 Moscow,  Russia }
   \\
   Alexei Ternov
   \\
   \small{\it Department of Theoretical Physics, Moscow Institute for Physics and Technology,}
   \\
   \small {\it 141700 Dolgoprudny, Russia }}

\date{}
\maketitle

\sloppy

\begin{abstract}
The claim of preprint hep-ph/0605114 that there is \ ``no neutrino spin light
because of photon dispersion in medium" is wrong.
\end{abstract}

In a series of our papers \cite{LobStudPLB03} -\cite{StuNPB05}, we have
proposed and studied in detail a new type of electromagnetic radiation of a
massive neutrino moving in the background matter which has been termed the \
``spin light of neutrino" ($SL\nu$) in matter. As it is well known (see, for
instance, \cite{Ginz60} and \cite{Shef75}), plasma is transparent for
electromagnetic radiation on frequencies  greater than the plasmon frequency.
In \cite{LobStudPLB03} -\cite{DvoGriStudIJMPD05}, we have shown from the
energy-momentum conservation law that a relativistic neutrino can emit the
$SL\nu$ photons with characteristic energy equals to a reasonable fraction of
the neutrino energy. So that if the neutrino energy much exceeds the plasmon
frequency (that is the case for a wide range of matter densities peculiar to
various astrophysical environments ) then the plasma influence on photons can
be neglected \cite{GriStuTerPLB_05, GriStuTerCOSMION04_hep_ph0502231}.

In the recent preprint hep-ph/0605114 \cite{KuznMikhhep-ph0605114}, it is
claimed that there is \ ``no neutrino spin light because of photon dispersion
in medium". This conclusion is based on uncorrect evaluation of the $SL\nu$
photon energy undertaken in \cite{KuznMikhhep-ph0605114} where the momentum
conservation law has not been accounted for. Thus, the mentioned above
statement of \cite{KuznMikhhep-ph0605114} is wrong.

\end{document}